# Electric field thermopower modulation analysis of an interfacial conducting layer formed between $Y_2O_3$ and rutile $TiO_2$

Taku Mizuno[1], Yuki Nagao[1], Akira Yoshikawa[1], Kunihito Koumoto[1], Takeharu Kato[2], Yuichi Ikuhara[2,3], and Hiromichi Ohta[1,4,a]

[1]*Graduate School of Engineering, Nagoya University, Furo–cho, Chikusa, Nagoya 464–8603, Japan*

[2]*Nanostructures Research Laboratory, Japan Fine Ceramics Center, Mutsuno, Atsuta, Nagoya 456–8587, Japan*

[3]*Institute of Engineering Innovation, The University of Tokyo, Bunkyo, Tokyo 113–8656, Japan*

[4]*PRESTO, Japan Science and Technology Agency, Sanban-cho, Tokyo 102–0075, Japan*

[a]Correspondence should be addressed H.O. (h-ohta@apchem.nagoya-u.ac.jp)

Electric field modulation analysis of thermopower ($S$) – carrier concentration ($n$) relation of a bilayer laminate structure composed of a 1.5-nm thick conducting layer, probably Ti$_n$O$_{2n-1}$ ($n$=2, 3, …) Magnéli phase, and rutile TiO$_2$ was performed. The results clearly showed that both the rutile TiO$_2$ and the thin interfacial layer contribute to carrier transport: the rutile TiO$_2$ bulk region (mobility $\mu$~0.03 cm$^2$V$^{-1}$s$^{-1}$) and the 1.5-nm thick interfacial layer ($\mu$~0.3 cm$^2$V$^{-1}$s$^{-1}$). The effective thickness of the interfacial layer, which was obtained from the $S$-$n$ relation, was below ~ 3 nm, which agrees well with that of the TEM observation (~1.5 nm), clearly showing that electric field modulation measurement of $S$-$n$ relation can effectively clarify the carrier transport properties of a bilayer laminate structure.

**I. INTRODUCTION**

Recently, field effect transistor (FET) structures on semiconductors have attracted attention to simultaneously clarify their carrier transport properties, *i.e.* carrier concentration (*n*), mobility (*µ*), and thermopower (*S*), of the two-dimensional (2D) channel, which can be formed at the gate insulator/semiconductor interface by applying a gate voltage ($V_g$).[1-11] Most material researchers usually measure the *µ–n* relation of the 2D channel.[1-2, 5] However, it is very difficult to extract the intrinsic carrier transport properties of the 2D channel using the *µ–n* relation because *µ* value is sensitive to the quality of the specimen, *i.e.*, structural defects and grain boundaries.

On the contrary, *S* value is rather insensitive to the specimen quality, and clearly depends on the energy derivative of the electronic density of states (DOS) at the Fermi energy ($E_F$), $\left[\frac{\partial DOS(E)}{\partial E}\right]_{E=E_F}$. Here, energy-dependent electrical conductivity $\sigma(E)$ is expressed as $\sigma(E) = n(E)e\mu(E)$ taken at the $E_F$ with $n(E)=DOS(E) f(E)$, where *f(E)* is the Fermi function, and *e* the electron charge. Hence, *S* is given by the following Mott equation,[12]

$$S = \frac{\pi^2}{3}\frac{k_B^2 T}{e}\left\{\frac{d[\ln(\sigma(E))]}{dE}\right\}_{E=E_F}.$$

$$= \frac{\pi^2}{3} \frac{k_B^2 T}{e} \left\{ \frac{1}{n} \cdot \frac{dn(E)}{dE} + \frac{1}{\mu} \cdot \frac{d\mu(E)}{dE} \right\}_{E=E_F}$$

Therefore, we aim at electric field modulation of $S$–$n$ relation to clarify intrinsic carrier transport properties of a semiconductor.

We have confirmed that the field modulation of $S$ method is applicable to thermoelectric materials using $SrTiO_3$[7]– and $KTaO_3$[8]–FETs. Furthermore, we have used this method to clarify the DOS of anatase $TiO_2$[9] and transparent amorphous oxide semiconductors, $InGaZnO_4$ and $In_2MgO_4$,[10] which are composed of both the parabolic shaped original DOS and a small non-parabolic shaped DOS such as a tail state. Thus, we have demonstrated the effectiveness of electric field modulation of $S$ using several monolayer materials.

To expand the applicability of the field modulation of $S$ method, we are currently focusing on clarifying the carrier transport properties of bilayer laminate materials, which are composed of two conducting layers A and B. The observable $S$ value can be expressed as $S = \dfrac{S_A \cdot \sigma_{2DA} + S_B \cdot \sigma_{2DB}}{\sigma_{2DA} + \sigma_{2DB}}$ where $\sigma_{2D}$ is the sheet conductance of the A or B layer.[13] Thus, the $\sigma_{2D}$ values of the A and B layers can be determined by analyzing the observed $S$ values. Hence, we have employed the field modulation of $S$ method to elucidate the carrier transports properties of a bilayer laminate material.

Herein we demonstrate that the electric field modulation of $S$ can effectively clarify the carrier transport properties of a bilayer laminate structure. We obtained a bilayer laminate with a 1.5-nm-thick interfacial conducting layer, probably $Ti_nO_{2n-1}$ ($n$=2, 3, …) Magnéli phase,[14,15] which was formed at the insulating $Y_2O_3$/$TiO_2$ heterointerface when we fabricated a polycrystalline $Y_2O_3$ thin film on a single crystalline rutile $TiO_2$ by pulsed laser deposition (PLD) at room temperature (RT). We then fabricated a FET on $Y_2O_3$/$TiO_2$ and measured the $S$-$n$ relation. The results clearly showed that both the original $TiO_2$ and the thin interfacial layer contribute to carrier transport: the rutile $TiO_2$ bulk region ($\mu$~0.03 $cm^2V^{-1}s^{-1}$) and the 1.5-nm thick interfacial layer ($\mu$~0.3 $cm^2V^{-1}s^{-1}$). The effective thickness of the interfacial layer, which was obtained from the $S$-$n$ relation, was below ~ 3 nm, which agrees well with that of the TEM observation.

## II. EXPERIMENTAL

First, a 200-nm-thick polycrystalline $Y_2O_3$ thin film was deposited on a (100) rutile $TiO_2$ single crystal plate, which was previously treated with a HF solution,[16] by PLD (KrF excimer laser, ~3 $Jcm^{-2}pulse^{-1}$, oxygen pressure 1 Pa) at RT. Figure 1(a) shows a cross-sectional high-resolution transmission electron microscopy image of the $Y_2O_3$/rutile $TiO_2$ interface region (HRTEM, TOPCON EM-002B, acceleration voltage of 200 kV). Although the rutile $TiO_2$ layer exhibited a uniform lattice [Fig. 1(a) bottom], $Y_2O_3$ had a polycrystalline nature [Fig. 1(a) top].

Additionally, a ~1.5-nm-thick interfacial layer was observed at the heterointerface. We confirmed that the interfacial layer was composed of Ti and O, and did not contain any Y related species by energy dispersive X-ray spectroscopy (EDS, data not shown) mapping of the heterointerface region. The atomic arrangement of the interfacial layer differed from that of the rutile layer, but the interfacial layer and rutile crystal were smoothly connected at the interface. We speculated that $Ti_nO_{2n-1}$ ($n$=2, 3, …), known as Magnéli, was formed at the top surface of the rutile crystal during $Y_2O_3$ film deposition. Oxygen-deficient $Y_2O_{3-\delta}$ was probably formed by the PLD process and the deposited $Y_2O_{3-\delta}$ film may extract oxide ions ($O^{2-}$) from the rutile $TiO_2$ substrate, resulting in the formation of Magnéli $Ti_nO_{2n-1}$.[14,15]

In order to clarify the carrier transport properties of the interfacial layer, we then fabricated a FET structure on $Y_2O_3$/$TiO_2$ using the $Y_2O_3$ layer as a gate insulator (dielectric permittivity, $\varepsilon_r$ =20[9,17]) [Fig. 1(b),(c)]. First, metallic Ti films (20-nm-thick), which were used as source (S) and drain (D) electrodes, were deposited by electron beam (EB, without substrate heating, base pressure ~$10^{-4}$ Pa) evaporation through a stencil mask onto a (100) rutile TiO$_2$ single crystal plate which was previously treated with a HF solution. Second, polycrystalline Y$_2$O$_3$ film (200-nm-thick) deposited through a stencil mask by PLD as described above, then the interfacial layer was introduced at Y$_2$O$_3$/TiO$_2$ interface. Finally, metallic Ti film (20-nm-thick), which was used as a gate (G) electrode, was deposited through a stencil mask by EB evaporation. Details of our FET fabrication process are described in the figure captions and elsewhere.[7-11]

The transistor characteristics of the resultant rutile TiO$_2$ based FETs were measured using a semiconductor device analyzer (B1500A, Agilent Technologies) at RT. The channel length ($L$) and width ($W$) were 200 and 400 μm, respectively. For thermopower ($S$) measurements, the FET was placed on two Peltier devices, which were used to introduce a temperature gradient between the S and D electrodes, as schematically shown in Fig. 1(b). Using manipulators, two thermocouples (T.C., type K, diameter: 150 μm, time constant ~25 ms, SHINNETSU Co.) were mechanically attached to both edges of the channel to measure the temperature difference ($\Delta T$) and the S and D probes measured the thermo-electromotive force ($\Delta V$) [Fig. 1(c)]. $S$-values were obtained from the slope of $\Delta V$–$\Delta T$ plots as shown in Fig. 1(d). It should be noted that we ignored the effect of the thermal voltage of the Ti contact pad because the $S$-value of metallic Ti (20 nm) contact is a few μVK$^{-1}$ at RT, far smaller than that of TiO$_2$. Further, to simplify the argument, we merely considered $\Delta T$ along the layers because $\Delta T$ in the perpendicular direction should be negligibly small.

**III. RESULTS AND DISCUSSION**

First, we confirmed the existence of the thin interfacial layer by analyzing the transistor characteristics; the electron transport properties of the interfacial layer overlap with that of the original rutile TiO$_2$. Figure 2 summarizes the FET characteristics of the resultant rutile TiO$_2$-based FET at RT. [(a) Typical drain current ($I_d$)–gate voltage ($V_g$), capacitance ($C$)–$V_g$, and $\sqrt{I_d}$–$V_g$ characteristics, (b) Field effect mobility ($\mu_{FE}$) and sheet carrier concentration $n_{2D(1)}$ and $n_{2D(2)}$]. A two-step $I_d$ increase was observed in the $I_d$–$V_g$ characteristic [Fig. 2(a)]. The $C$–$V_g$ curve (inset) also exhibited a similar increase. $C$ became saturated (93 nFcm$^{-2}$) near $V_g$=+2 V. The two threshold voltages ($V_{gth(1)}$ and $V_{gth(2)}$) were estimated from the $\sqrt{I_d}$–$V_g$ characteristic as $V_{gth(1)}$= –14.7 V and $V_{gth(2)}$= –6.8 V. The $V_{gth}$ values corresponded to the bending points in the $I_d$–$V_g$ and $C$–$V_g$ curves.

Moreover, field effect mobility ($\mu_{FE}$) also exhibited a two-step increase [Fig. 2(b)]. The $\mu_{FE}$ values were calculated from $\mu_{FE} = g_m[(W/L)C_{sat} \cdot V_d]^{-1}$ where $g_m$ is the transconductance $\partial I_d/\partial V_g$ and $C_{sat}$ is the saturated capacitance value (93 nFcm$^{-2}$). $\mu_{FE}$ reached ~0.03 cm$^2$V$^{-1}$s$^{-1}$ when $V_g \approx V_{gth(2)}$, which corresponds well with the value reported by Katayama *et al.*[1] (~0.05 cm$^2$V$^{-1}$s$^{-1}$). However, it reached ~0.3 cm$^2$V$^{-1}$s$^{-1}$ when $V_g > V_{gth(2)}$, which corresponds well with the room temperature Hall mobility of the Ti$_2$O$_3$ epitaxial film (~0.5 cm$^2$V$^{-1}$s$^{-1}$).[18] These results clearly indicate that the FET characteristics can be divided at $V_{gth(2)}$.

Next, we calculated the sheet carrier concentration ($n_{2D}$), which can be obtained from $n_{2D} = C(V_g - V_{gth})$ using both $V_{gth}$ values, $V_{gth(1)}$ and $V_{gth(2)}$, as shown in Fig. 2(a). Furthermore, we observed a clear pinch-off and current saturation in $I_d$ [Figs. 3(a),(b)]. For the lower $V_g$ ($V_g \leq -8$ V), the pinch-off $V_d$ (dotted line) corresponded the $V_g - V_{gth(1)}$, whereas the pinch-off $V_d$ corresponded to $V_g - V_{gth(2)}$ ($V_g \geq -6$ V). These results clearly suggest that two different conducting layers, the rutile TiO$_2$ bulk region ($n_{2D(1)}$, $\mu_{FE}$ ~0.03 cm$^2$V$^{-1}$s$^{-1}$) and the 1.5-nm thick interfacial layer ($n_{2D(2)}$, $\mu_{FE}$ ~0.3 cm$^2$V$^{-1}$s$^{-1}$), contribute to the operation of this FET.

Figure 4(a) shows the field modulated $S$ of the TiO$_2$–FET at RT. The relationship between $S$ and the bulk carrier concentration ($n_{3D}$) for bulk single crystal rutile TiO$_2$[19] is also shown in the inset together with that for a 0.5 wt.% Nb-doped rutile TiO$_2$ single crystal (SHINKOSHA Co.). The $S$ values were always negative as shown in Fig. 1(d), indicating *n*-type conductivity of the channels. Unfortunately, in the region of $V_g < V_{gth(2)}$, any reliable $S$-values were not obtained because the channel is highly resistive. In the region of $V_g > V_{gth(2)}$, the $|S|$ value gradually decreased from 0.7 to 0.5 mVK$^{-1}$ as $V_g$ increased due to an increase of $n_{3D}$ in the channel from $10^{18}$ to $10^{19}$ cm$^{-3}$ as plotted in Fig. 4(b).

Here we estimate the thickness of the interfacial layer using the observed $S$-values as follows. The observed $S$-values ($S_{obsd.}$) were expressed as,

$$S_{obsd.} = \frac{S_{Int.} \cdot \sigma_{2D\,Int} + S_{Rutile} \cdot \sigma_{2D\,Rutile}}{\sigma_{2D\,Int} + \sigma_{2D\,Rutile}}$$

, where $S_{Int.}$, $S_{Rutile}$, $\sigma_{2D\,Int.}$ and $\sigma_{2D\,Rutile}$ are $S$ of the interfacial layer, rutile, $\sigma_{2D}$ of the interfacial layer and rutile, respectively. In the region of $V_g > V_{gth(2)}$, since $\sigma_{2D\,Int.}$ is an order magnitude larger than $\sigma_{2D\,Rutile}$, $S_{Int.}$ dominantly contributes to $S_{obsd}$ as,

$$S_{obsd.} \approx \frac{S_{Int.} \cdot \sigma_{2D\,Int}}{\sigma_{2D\,Int} + \sigma_{2D\,Rutile}} \quad (\sigma_{2D\,Int} >> \sigma_{2D\,Rutile})$$

Therefore, we roughly estimated the thickness of the interfacial layer as $n_{2D(2)}/n_{3D}$ as shown in Fig. 4(b). The calculated $n_{2D(2)}/n_{3D}$ values were ~3 nm, which agree well with the interfacial layer thickness (~1.5 nm). These results clearly show that thermopower measurement using a FET structure can clarify the carrier transport properties of bilayer laminate structure.

**IV. SUMMARY**

In summary, we have demonstrated that the electric field modulation of $S$ can effectively clarify the carrier transport properties of a bilayer laminate structure composed of a 1.5-nm thick interfacial conducting layer and rutile $TiO_2$ single crystal. We fabricated an FET structure on the bilayer laminate and measured the $S$-$n$ relation at RT. The results clearly showed that both the original $TiO_2$ and the thin interfacial layer contribute to carrier transport: the rutile $TiO_2$ bulk region ($\mu \sim 0.03$ $cm^2V^{-1}s^{-1}$) and the 1.5-nm thick interfacial layer ($\mu \sim 0.3$ $cm^2V^{-1}s^{-1}$). The effective thickness of the interfacial layer, which was obtained from the $S$-$n$ relation, was below $\sim 3$ nm, which agrees well with that of the TEM observation. These results clearly show that thermopower measurement using a FET structure can clarify the carrier transport properties of bilayer laminate structure.

**ACKNOWLEDGEMENTS**

This work was partly supported by the Ministry of Education, Culture, Sports, Science and Technology (22360271, 22015009).

**References**

1. M. Katayama, S. Ikesaka, J. Kuwano, Y. Yamamoto, H. Koinuma, and Y. Matsumoto, *Appl. Phys. Lett*. **89**, 242103 (2006).

2. H. Nakamura, H. Takagi, I. H. Inoue, Y. Takahashi, T. Hasegawa, and Y. Tokura, *Appl. Phys. Lett*. **89**, 133504 (2006).

3. A. Mühlenen, N. Errien, M. Schaer, M-N. Bussac, and L. Zuppiroli, *Phys. Rev. B* **75**, 115338 (2007).

4. K. P. Pernstich, B. Rössner, and B. Batlogg, *Nature Mater*. **7**, 321 (2008).

5. K. Ueno, S. Nakamura, H. Shimotani, A. Ohtomo, N. Kimura, T. Nojima, H. Aoki, Y. Iwasa, and M. Kawasaki, *Nature Mater*. **7**, 855 (2008).

6. W. Liang, A. I. Hochbaum, M. Fardy, O. Rabin, M. Zhang, and P. Yang, *Nano Lett*. **9**, 1689 (2009).

7. H. Ohta, Y. Masuoka, R. Asahi, T. Kato, Y. Ikuhara, K. Nomura, and H. Hosono, *Appl. Phys. Lett*. **95**, 113505 (2009).

8. A. Yoshikawa, K. Uchida, K. Koumoto, T. Kato, Y. Ikuhara, and H. Ohta, *Appl. Phys. Express* **2**, 121103 (2009).

9. Y. Nagao, A. Yoshikawa, K. Koumoto, T. Kato, Y. Ikuhara, and H. Ohta, *Appl. Phys. Lett*. **97**, 172112 (2010).

10. H. Koide, Y. Nagao, K. Koumoto, Y. Takasaki, T. Umemura, T. Kato, Y. Ikuhara, and H. Ohta, *Appl. Phys. Lett*. **97**, 182105 (2010).

11. H. Ohta, Y. Sato, T. Kato, S-W. Kim, K. Nomura, Y. Ikuhara, and H. Hosono, *Nature Commun*. **1**, 118 (2010).

12. M. Cutler, and N. F. Mott, *Phys. Rev*. **181**, 1336 (1969).

13. T. Koga, S. B. Cronin, M. S. Dresselhaus, J. L. Liu, and K. L. Wang, *Appl. Phys. Lett*. **77**, 1490 (2000).

14. S. Andersson, and A. Magnéli, *Naturwissenschaften* **43**, 495 (1956).

15. M. Marezio, and P. D. Dernier, *J. Solid State Chem*. **3**, 340 (1971).

16. Y. Yamamoto, K. Nakajima, T Ohsawa, Y. Matsumoto, and H. Koinuma, *Jpn. J. Appl. Phys*. **44**, L511 (2005).

17. We measured the dielectric permittivity of the $Y_2O_3$ film (~20), which was similar to the reported value [~18: J. Kwo, M. Hong, A. R. Kortan, K. L. Queeney, Y. J. Chabal, R. L. Opila, D. A. Muller, S. N. G. Chu, B. J. Sapjeta, T. S. Lay, J. P. Mannaerts, T. Boone, H. W. Krautter, J. J. Krajewski, A. M. Sergnt, and J. M. Rosamilia, *J. Appl. Phys*. **89**, 3920 (2001)].

18. We measured $\mu_{Hall}$ value of a $Ti_2O_3$ ($Ti_nO_{2n-1}$, $n=2$) thin film at RT, which was heteroepitaxially grown on (0001) face of $\alpha$-$Al_2O_3$ substrate by PLD at the substrate temperature of 800°C in an oxygen atmosphere (~$10^{-5}$ Pa).

19. M. Itakura, N. Niizeki, H. Toyoda and H. Iwasaki, *Jpn. J. Appl. Phys*. **6**, 311 (1967).

**Figure captions**

**Fig. 1 (Color online).** Single crystalline rutile $TiO_2$-based FET structure. (a) Cross-sectional HRTEM image of the $Y_2O_3/TiO_2$ interface region. Bottom part clearly shows a rutile crystalline lattice, whereas the top part shows a randomly oriented $Y_2O_3$ crystal. Additionally, a ~1.5-nm thick interfacial layer is seen at the heterointerface. (b) Schematic device structure and (c) optical micrograph of the FET fabricated on a (100) rutile $TiO_2$ single crystal (0.5 mm$^t$). S, D, and G electrodes are 20-nm thick Ti films, and the G insulator ($\varepsilon_r$=20[9, 17]) is a 200-nm thick polycrystalline $Y_2O_3$ film. The channel direction is along with [001] direction. $L$ and $W$ are 200 and 400 μm, respectively. For the thermopower measurements, the FET is placed on two Peltier devices, which are used to introduce a temperature gradient between the S and D electrodes. Two T.C. (type K) are attached on both edges of the channel. (d) $\Delta V$–$\Delta T$ plots of the channel with +15 V of G voltage ($V_g$) applied is shown as an example. The values of $S$ were obtained from the slope of $\Delta V$–$\Delta T$ plots with various $V_g$.

**Fig. 2 (Color online).** FET characteristics of the rutile $TiO_2$-based FET. (a)

Typical $I_d$–$V_g$ ($V_d$ = +1 V) and $C$–$V_g$ (Frequency, $f$ =20 Hz) characteristics. $C$ is saturated (93 nFcm$^{-2}$) near $V_g$=+2 V. The $\sqrt{I_d}$–$V_g$ characteristic curve is also shown. Two $V_{gth}$ are seen ($V_{gth(1)}$ = –14.7 V, $V_{gth(2)}$ = –6.8 V). (b) $\mu_{FE}$ and $n_{2D}$. $n_{2D}$ values are estimated based on the standard field effect theory using $V_{gth(1)}$ and $V_{gth(2)}$. Two step increase of $\mu_{FE}$ is observed. $\mu_{FE}$ reaches ~0.03 cm$^2$V$^{-1}$s$^{-1}$ when $V_g \approx V_{gth(2)}$, but it reaches ~0.3 cm$^2$V$^{-1}$s$^{-1}$ when $V_g > V_{gth(2)}$, which is an order magnitude larger than the rutile-based FET (literature value ~0.05 cm$^2$V$^{-1}$s$^{-1}$).

**Fig. 3 (Color online).** $I_d$–$V_d$ characteristics. Pinch-off behaviors at (a) $V_g$–$V_{gth(1)}$ for $V_g \leq$ –8 V and (b) $V_g$–$V_{gth(2)}$ for $V_g \geq$ –6 V are seen.

**Fig. 4 (Color online).** Simple estimation of the effective channel thickness. (a) Thermopower ($S$) modulation of the TiO$_2$–FET at RT. |$S$| value gradually decreases from 0.7 to 0.5 mVK$^{-1}$ as $V_g$ increases ($V_g > V_{gth(2)}$). (Inset) Relationship between |$S$| and bulk carrier concentration ($n_{3D}$) for bulk single crystal TiO$_2$ (Ref. 19) and 0.5 wt.% Nb-doped TiO$_2$ single crystal". Fitted curve well reproduces the observed |$S$| data plots. (b) $n_{2D(2)}/n_{3D}$, which should roughly indicate the effective channel thickness, are obtained using the relationship

between the $n_{2D(2)}$ and $n_{3D}$. $n_{2D(2)}/n_{3D}$ value of ~3 nm corresponds well with the interfacial layer thickness (~1.5 nm).

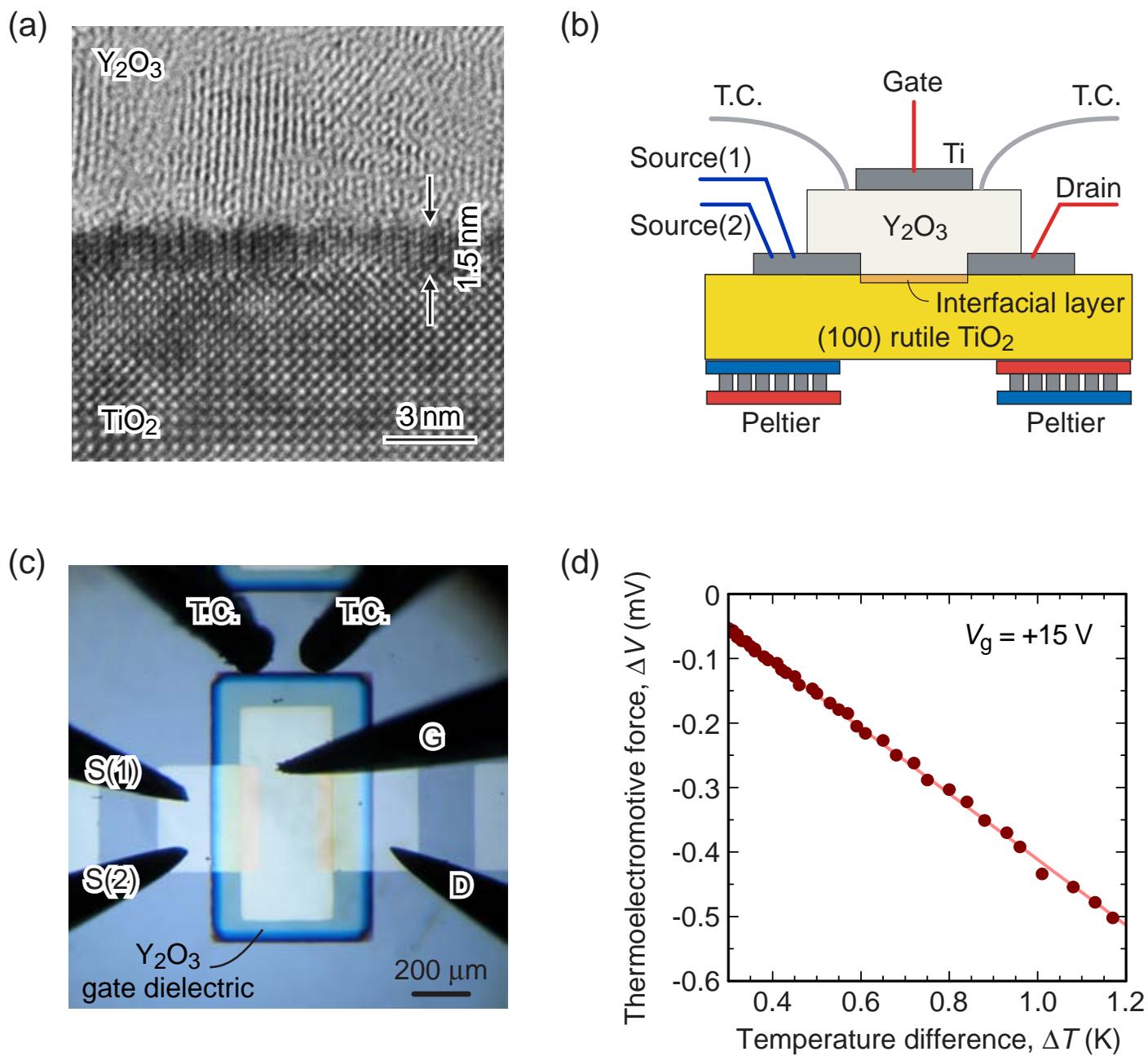

Fig. 1

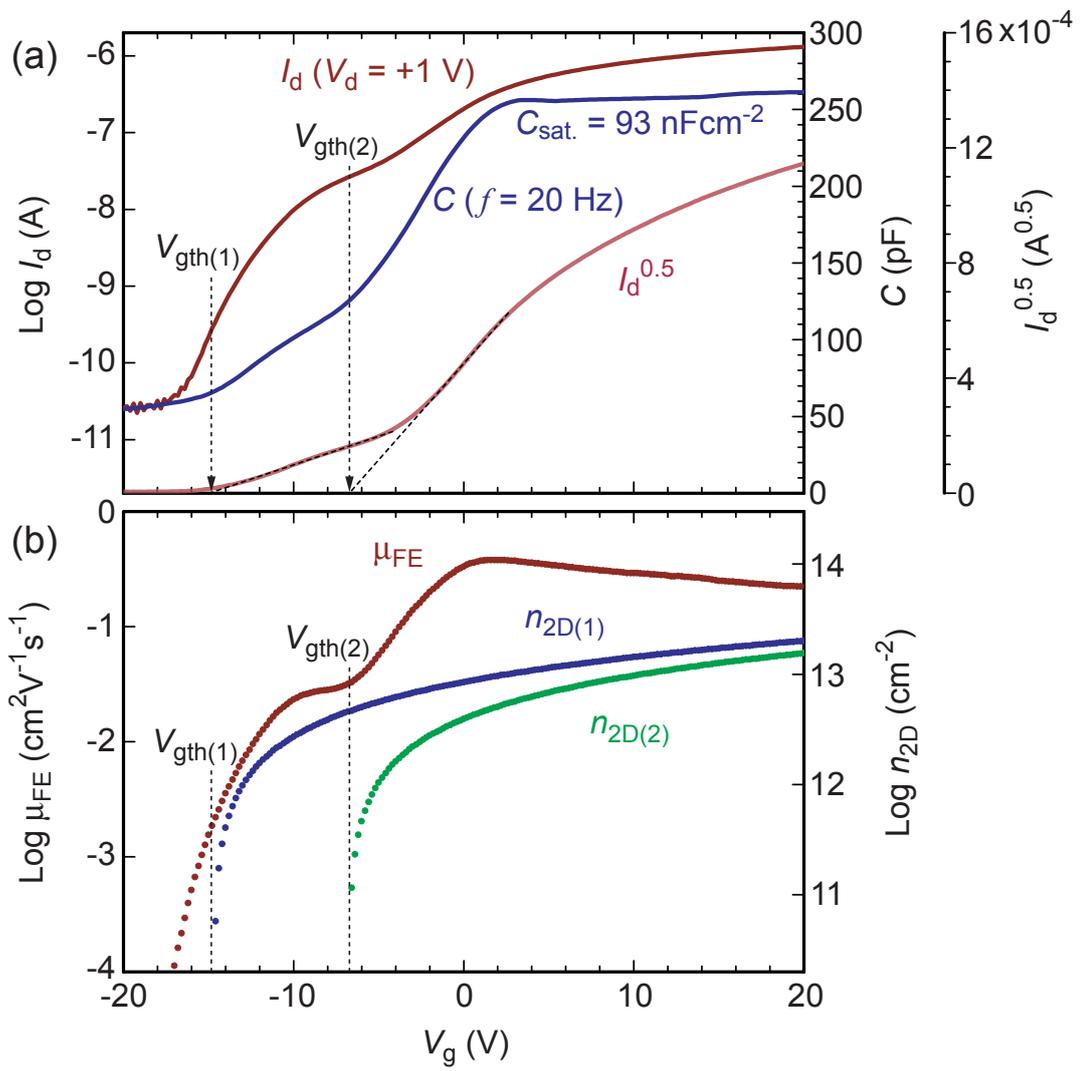

Fig. 2

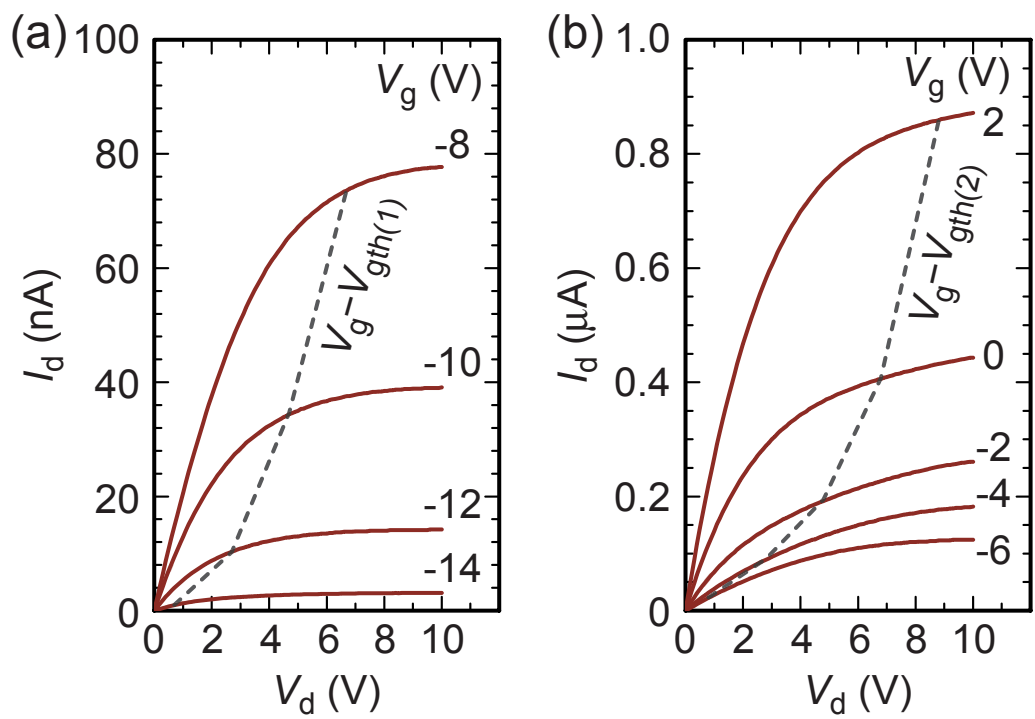

Fig. 3

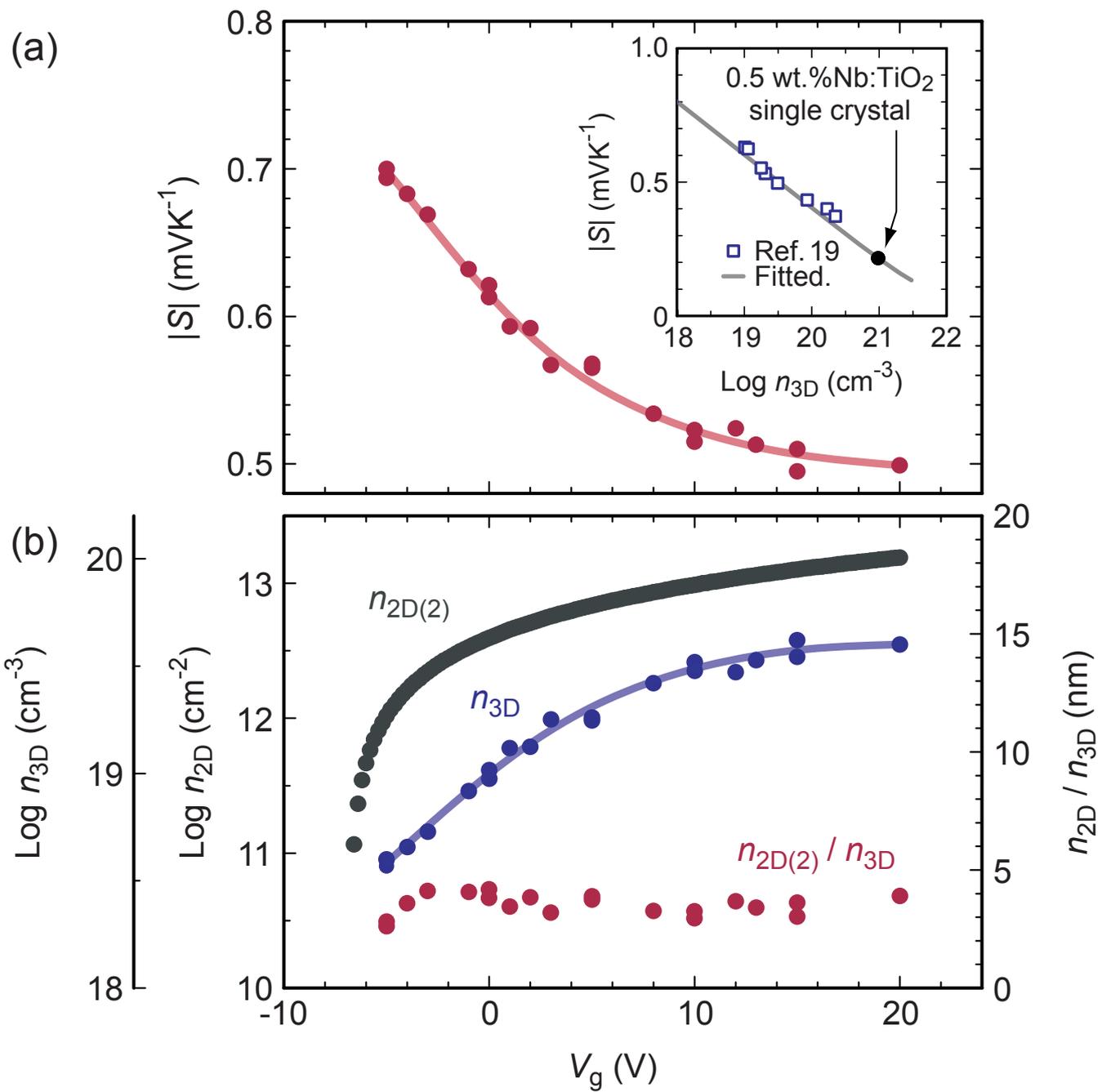

Fig. 4